\documentclass{article}
\usepackage[utf8]{inputenc}
\usepackage{authblk}
\usepackage{setspace}
\usepackage[margin=1.25in]{geometry}
\usepackage{graphicx}
\graphicspath{ {./manuscript/} }
\usepackage{subcaption}
\usepackage{amsmath,amsfonts,amssymb,physics,dirtytalk}
\usepackage{lineno}
\usepackage{hyperref}
\usepackage{comment}
\usepackage{siunitx}

\newcommand{\fzu}{\ $\si{erg\ cm^{-2}\ s^{-1}}$}

\title{Performance Evaluation of a silicon-based 6U Cubesat detector for soft $\gamma$-ray astronomy}

\author[1$\dag$]{Rishank Diwan}
\author[1$\dag$]{Kees de Kuijper}
\author[1,3*]{Partha Sarathi Pal}
\author[1]{Andreas Ritter}
\author[2]{Pablo M. Saz Parkinson}
\author[1]{Andy~C.~T.~Kong}
\author[1]{Quentin A. Parker}

\affil[1]{Laboratory for Space Research, Faculty of Science, The University of Hong Kong, 405B Cyberport 4, 100 Cyberport Road, Cyberport, Hong Kong, China}
\affil[2]{Santa Cruz Institute for Particle Physics and Department of Physics, University of California, Santa Cruz, CA, 95064, USA}
\affil[3]{Institute of Astronomy Space and Earth Science, P177, CIT Road, Scheme-7M, Kolkata, 700054, West Bengal, India}
\affil[*]{Address correspondence to: parthasarathi.pal@gmail.com}
\affil[$\dag$]{These authors contributed equally to this work.}

\date{}

\begin{document}

\maketitle

\begin{abstract}
The observation of the low-energy $\gamma$-ray (0.1-30 MeV) sky has been significantly limited since the COMPTEL instrument was decommissioned aboard the Compton Gamma-ray Observer (CGRO) satellite in 2000. The exploration of $\gamma$-ray photons within this energy band, often referred to as the \say{MeV gap}, is crucial to address numerous unresolved mysteries in high-energy and multi-messenger astrophysics. Although several large MeV $\gamma$-ray missions have been proposed (e.g., e-ASTROGAM, AMEGO, COSI), most of these are in the planning phase, with launches not expected until the next decade, at the earliest. Recently, there has been a surge in proposed CubeSat missions as cost-effective and rapidly implementable \say{pathfinder} alternatives. A MeV CubeSat payload dedicated to $\gamma$-ray astronomy could serve as a valuable demonstrator for large-scale future MeV payloads. This paper proposes a $\gamma$-ray payload design with a Silicon-based tracker and a Ceasium-Iodide-based calorimeter. We report the results of a simulation study to assess the performance of this payload concept and compare the results with those of previous $\gamma$-ray instruments. As part of the performance assessment and comparison, we show that with our proposed payload design, a sensitivity better than IBIS can be achieved for energies between 0.1 and 10 MeV, and for energies up to around 1 MeV, the achieved sensitivity is comparable to COMPTEL, therefore opening up a window towards cost-effective observational astronomy with comparable performance to past missions.
\end{abstract}

\section{Introduction}
\label{introduction}
Gamma-ray astronomy has matured significantly over the course of the last decade and a half, leading to the exploration of the high-energy Universe with unprecedented sensitivity. The $Fermi$ Large Area Telescope (LAT) \cite{Atwood09} has continuously observed the $\gamma$-ray sky between a few tens of MeV up to $>$300 GeV since its launch in 2008. During this time, the \textit{Fermi}-LAT has observed gamma rays from numerous source classes (e.g. the Galactic Center \cite{Ackermann_2017,PhysRevD.103.063029}, Pulsars \cite{smith2023fermi}, Active Galactic Nuclei (AGN) \cite{Ajello_2020}, Blazars \cite{Kerby_2023,2014ApJ...780...73A}, Gamma-Ray Bursts (GRBs) \cite{Ajello_2019} etc.). Indeed, the analysis of $Fermi$ data has also led to several unexpected astronomical discoveries in $\gamma$-ray astronomy (e.g. the $Fermi$ Bubbles \cite{Su_2010, ackermann2014spectrum, galaxies6010029}, a population of radio-quiet $\gamma$-ray pulsars \cite{Abdo09}, classical $\gamma$-ray novae \cite{Ackermann14}, numerous extreme blazars/AGN \cite{Ackermann_2015}, high-energy solar flares \cite{Ajello_2021}, $\gamma$-ray flashes in thunderstorms \cite{https://doi.org/10.1029/2017JA024837}, etc.), as well as major breakthroughs in multi-messenger astronomy \cite{Abbott17, TXS0506, PhysRevD.103.123005}.

Despite these significant successes in the GeV $\gamma$-ray range, the low-energy (MeV) $\gamma$-ray sky remains poorly observed, especially between 0.1 and 30 MeV. Since the de-orbiting of the Imaging Compton Telescope (COMPTEL) \cite{Schonfelder1993, ryan23} on board the Compton Gamma-ray Observatory (CGRO) in 2000, there have been no specialized missions to study this energy range with good sensitivity, thus creating an observational gap in MeV energy range, sometimes referred to as the \say{MeV gap}. After more than two decades, an in-depth exploration of this energy range is critical for improving the understanding of astrophysical sources and the burgeoning field of \say{multi-messenger} astronomy. The study of GRBs, MeV blazars, MeV pulsars, and the MeV background are all topics of significant interest to the $\gamma$-ray community. Moreover, more profound studies of the 511 keV emission line resulting from electron-positron annihilation would potentially enhance the understanding of systems like Type Ia supernovae (SNIa) \cite{2018MNRAS.480.1393C} and Galactic compact objects in general \cite{2011A&A...531A..56D}. 

Blazars are the most numerous class of high-energy $\gamma$-ray sources. To better understand the evolution of blazar Spectral Energy Distributions (SEDs), it is crucial to bridge the MeV gap between X-ray and $\gamma$-ray observations \cite{2017MNRAS.469..255G}. Many blazars are expected to exhibit a peak around a few MeV in their SEDs, a spectral region currently inaccessible due to the MeV gap \cite{Ajello_2016, Marcotulli_2020}. Incorporating MeV photons into blazar SED modeling is essential, particularly given the current expectation of Blazars as potential neutrino sources, making them prime targets in the current era of multi-messenger astronomy \cite{boettcher2022multiwavelength}. However, effective neutrino observations require simultaneous $\gamma$-ray observations for counterpart identification. At present, no definitive correlations have been established between neutrinos and $\gamma$-ray photons, although marginal evidence has been found ~\cite{TXS0506}. Thus, a targeted search for counterparts in the MeV energy band is essential for a better understanding of future neutrino events \cite{murase2016hidden}. 

Other topics of interest to the $\gamma$-ray astronomy community include the evolution of the 1.8 MeV Galactic Aluminium-26 emission line, which serves as a tracer of the nucleosynthesis of radioactive elements in the Galactic Center region, making it an invaluable tool for the study of this complex region \cite{Beechert_2022}. In addition, the diffuse MeV background remains an open question and the subject of intense research. While it is hypothesized to comprise MeV photons from cosmic rays, AGN, SNIa, star-forming galaxies, etc. \cite{Ruiz-Lapuente_2016}, only new observations hold the potential to constrain current models effectively.  

At present, several large-scale $\gamma$-ray telescopes sensitive to the MeV energy range are under development, such as e-ASTROGAM \cite{2018JHEAp..19....1D}, AMEGO \cite{AMEGOref}, or COSI \cite{tomsick2019compton, Tomsick:2021H5}. Of these, only COSI is currently fully funded and expected to launch within the coming decade. Within CubeSat configurations, several projects are in various stages of development \cite{Racusin17, Wen21, Bloser2022}, including a Double-Sided Silicon Strip Detector (DSSD) payload design, with the added capability of detecting polarisation from $\gamma$-ray sources within the MeV range \cite{Yang_2020}.
Recently, besides DSSD, different technologies such as CZT crystal detectors and Scintillators are being used for MeV observations \cite{2022JInst..17P8004L, 2022JCAP...08..013L, mass2023, 2024ExA....57...16D, gagg2024, compair2024, BOLOTNIKOV2024169328}. With CZT crystal detectors, only untracked analysis of $\gamma$-ray photons is possible. In contrast, silicon tracker-based detectors can do both tracked and untracked analysis, giving better source location reconstruction accuracy than CZT-based detectors.
This paper presents the design and performance of a compact 3U detector and a further scaled-down 2U version tailored for a CubeSat mission, enabling observations within the desired energy range. Such a CubeSat mission could serve as a pathfinder, providing valuable insights into the performance expectations of larger-scale MeV telescopes.

\section{Payload simulation and performance calculation}
\subsection{Detector design}
\label{det_design}
\begin{figure*}
    \centering
    \begin{subfigure}{0.485\textwidth}
        \includegraphics[width=\textwidth, trim={2.5cm 4.5cm 3cm 3cm}, clip]{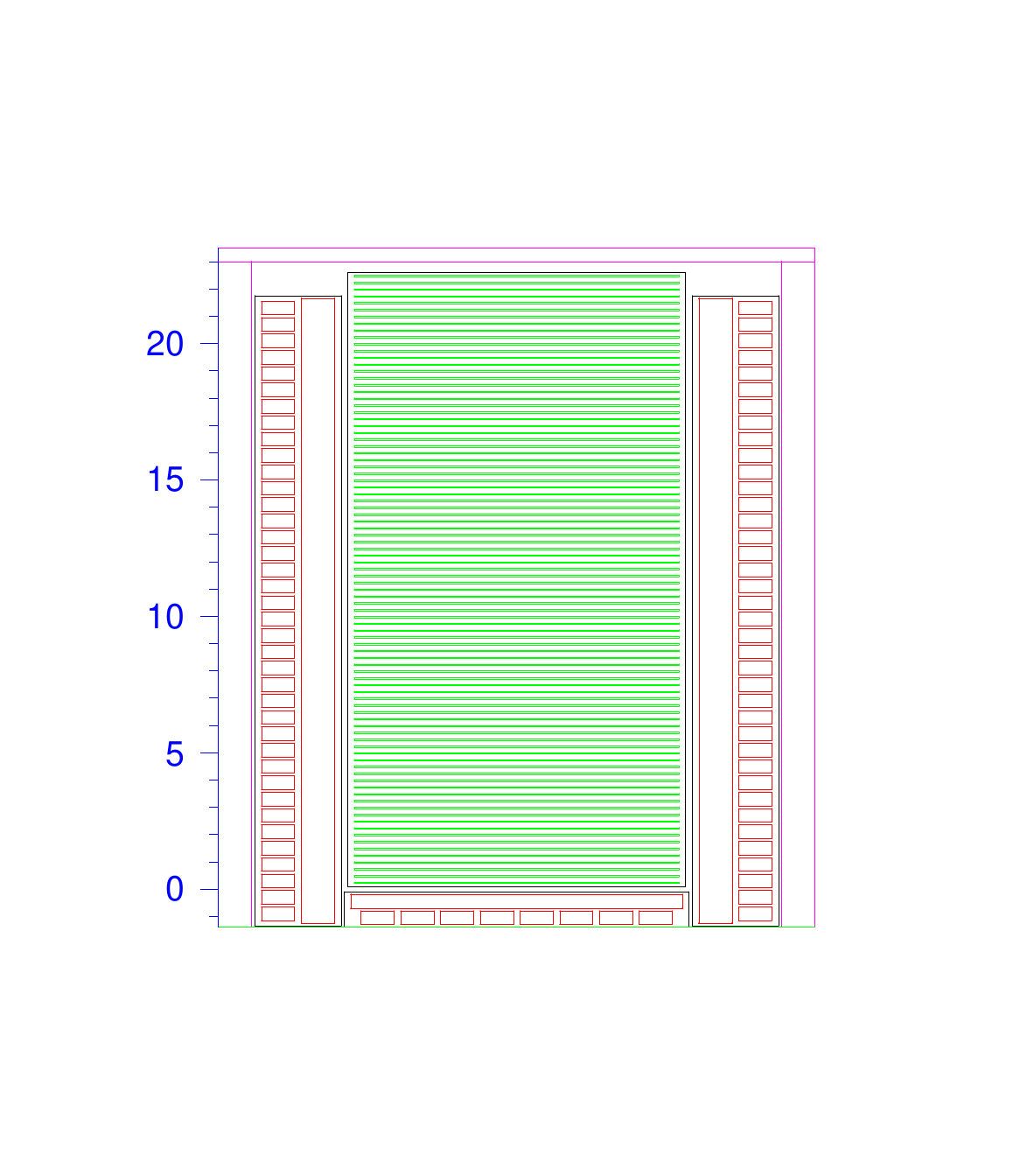}
        \caption{Side View}
        \label{side_3u}
    \end{subfigure}
    \begin{subfigure}{0.485\textwidth}
        \includegraphics[width=\textwidth, trim={3cm 2cm 3cm 3cm}, clip]{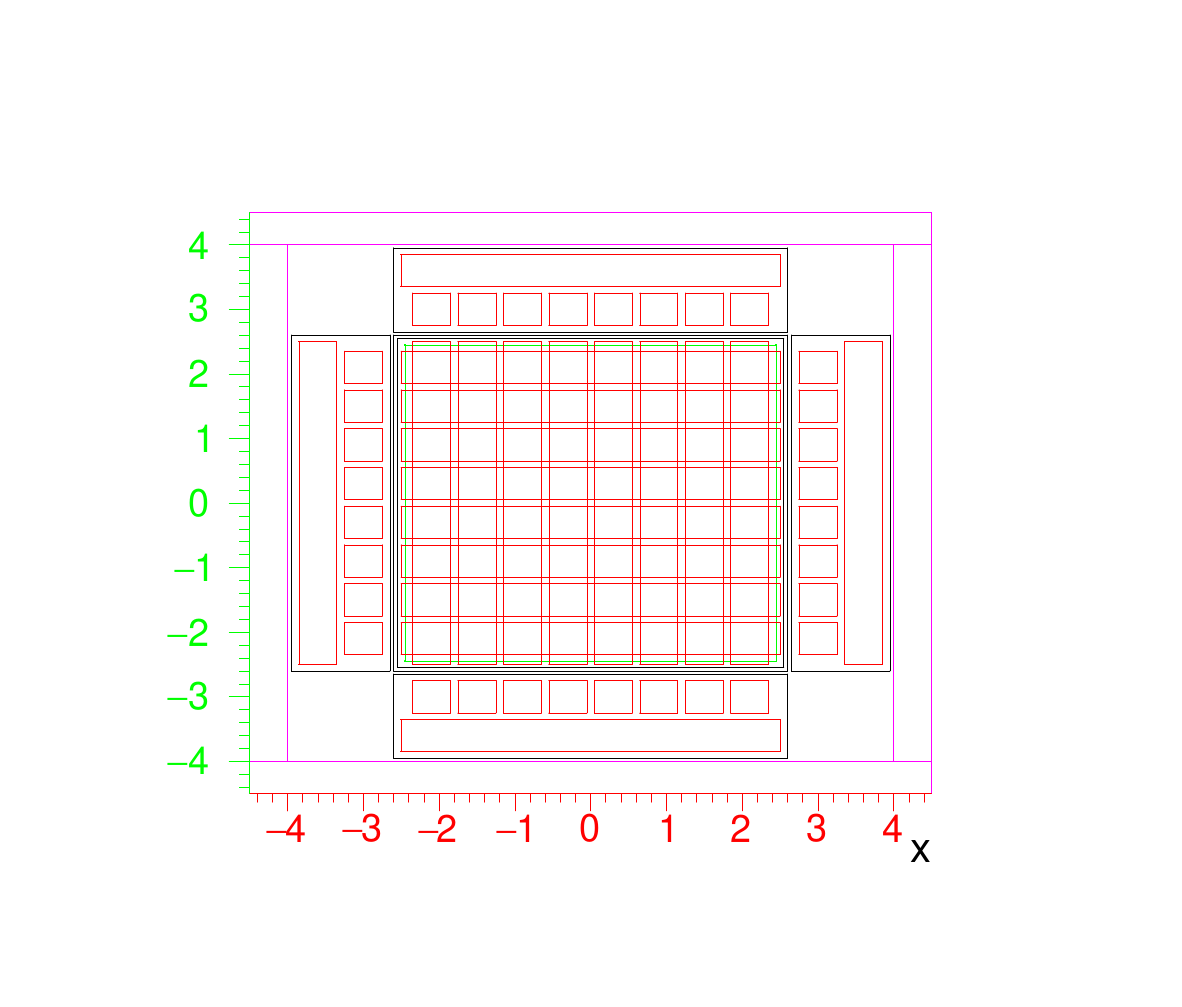}
        \caption{Top View}
        \label{top_3u}
    \end{subfigure}
    \caption{3U detector design. All units are in cm. Included detector components are the tracker (green), calorimeter (red), and anti-coincidence detector (magenta).}
    \label{det_3u}
\end{figure*}

Our MeV detector design is based on stacking a series of CubeSat micro-satellites. A CubeSat is a class of {\it nano satellite}, adhering to standardized size and weight requirements. A single unit (1U)\footnotemark[1] \footnotetext[1]{\url{https://www.nasa.gov/content/what-are-smallsats-and-cubesats}} must adhere to specifications of $10 \times 10 \times 10$ $\text{cm}^3$ and 1.33 kg. Multiple CubeSat units can be combined to form 1U, 1.5U, 2U, 3U and 6U, with possible extensions to 12U, 16U and even 24U. The MeV payload proposed in this paper draws inspiration from previous CubeSat designs~\cite{2019AJ....158...42R, 2017AJ....153..237L}, extending it to a 3U payload ($10 \times 10 \times 30$ $\text{cm}^3$) and a 2U payload ($10 \times 10 \times 20$ $\text{cm}^3$). Figure~\ref{det_3u} shows the side and top views of the 3U detector assembly, comprising three components: (i) Silicon Tracker, (ii) CsI Calorimeter, and (iii) Anti-Coincidence Detector (ACD). The entire CubeSat is expected to be roughly 6U in size ($10 \times 20 \times 30$ $\text{cm}^3$), with the non-payload volume of the satellite reserved for electronics and other commercial-off-the-shelf (COTS) components, depending on the constraints imposed by the final choice of payload (e.g., cost, dimensions).

\subsubsection{Silicon Tracker}
\label{tracker_section}
A crucial component of the MeV-CubeSat payload is the silicon tracker, as is the case in the $Fermi$ LAT $\gamma$-ray telescope \cite{2011JInst...6C2043B, 2013iwvd.confE..39B, 2023NIMPA104968080M}, which is behind the primary inspiration for the design choices. For the detector design, we opt for a double-sided cross-strip silicon detector (DSSD) to provide the tracking information for $\gamma$-rays, electrons, and positrons generated during Compton scattering and pair production. Our payload uses 90 (60) layers of Silicon distributed in 90 (60) layers of DSSD in the 3U (2U) tracker, with a $2$ mm tracker pitch. Precise tracking details are vital for reconstructing the path of incident $\gamma$-ray photons from the observed point source. In addition, tracking information yields the scatter plane deviation (SPD) angle and angular resolution measure (ARM) for Compton events, as well as the point spread function (PSF) value for the pair-production events. Furthermore, these values are crucial in calculating the payload's \textit{Sensitivity} $F_z$. It is important to note that the silicon tracker constitutes around 90$\%$ of the total payload budget.

\subsubsection{Calorimeter}
\label{cal_section}
The calorimeter is another vital component of the MeV-CubeSat payload design. The current configuration consists of two sets of cuboid-shaped caesium-iodide (CsI) scintillation crystals measuring 5 cm $\times$ 0.5 cm $\times$ 0.5 cm and 22.9 cm (15.2 cm) $\times$ 0.5 cm $\times$ 0.5 cm for the 3U and 2U configurations, respectively. These crystals are instrumental in measuring the energy of the incident and scattered $\gamma$-rays. The calorimeter design envelops the Silicon tracker from five sides, employing two layers of small-size CsI crystals arranged in an 8 $\times$ 8 orthogonal configuration. In contrast, on the lateral side of the payload, two layers consisting of a small calorimeter and a long calorimeter are again arranged orthogonally. The entire payload requires a total of 168 (116) small and 32 long crystals.

\subsubsection{Anti Coincidence Detector}
\label{acd_section}
Lastly, an anti-coincidence detector (ACD) made from plastic scintillation material shields the silicon tracker and the calorimeter. The ACD is to veto (reject) 99.99\% of background radiation originating from charged particles, such as the Earth's albedo or cosmic rays. This type of background radiation is prevalent in low-earth orbit (LEO) at an altitude of $\sim500$ km, which is the optimal orbit for this CubeSat mission. The \say{MeV gap} contains various background components that have been observed and modeled using data from instruments like $Fermi$-LAT \cite{FermiLATref} and INTEGRAL \cite{INTEGRALref}.

\subsection{Simulations and performance calculation} 
The detector design and its performance are simulated using \textit{MEGAlib}\footnotemark[2] \footnotetext[2]{\url{https://megalibtoolkit.com/home.html}} \cite{2006NewAR..50..629Z}, a software package based on Geant4 \cite{Agostinelli, Allison2006, Allison2016} and ROOT \cite{BrunRad}.
The detector design described in Sec.~\ref{det_design} is shown in Fig~\ref{det_3u}, which is simulated using the \textit{Geomega} module in MEGAlib. The simulation focuses only on detector components responsible for interactions with $\gamma$-ray photons (i.e., silicon tracker, CsI calorimeter, and ACD), excluding the electronic components and noise for this study.

The effective area, energy resolution, and angular resolution for the simulations were performed using monochromatic point sources with energies $0.1 \leq E \leq 6$ MeV, simultaneously considering albedo photon- and cosmic-ray background radiation with the following energy spectra:
\begin{equation}
\label{Bkgspectrum}
        \Big(\frac{dN}{dE}\Big)_\text{albedo} = c_1 E^{-\Gamma_1},\
        \Big(\frac{dN}{dE}\Big)_\text{cosmic} = c_2 E^{-\Gamma_2},
\end{equation}
where $c_1 = 0.329\ \si{ph\ cm^{-2}\ s^{-1}\ keV^{-1}\ sr^{-1}}$, $\Gamma_1 = 1.34$ (for 868 keV - 20 MeV); \\ $c_1 = 858\ \si{ph\ cm^{-2}\ s^{-1}\ keV^{-1}\ sr^{-1}}$, $\Gamma_1 = 2.12$ (for 20 MeV - 97.4 GeV); and \\ $c_2 = 11.3\ \si{ph\ cm^{-2}\ s^{-1}\ keV^{-1}\ sr^{-1}}$, $\Gamma_2 = 2.15$ (for 868 keV - 107 GeV). \\ This background model was obtained by fitting background spectrum data provided from the MEGAlib GitHub repository\footnotemark[3] \footnotetext[3]{\url{https://github.com/zoglauer/megalib/tree/main/resource/examples/cosima/source}}.
Furthermore, the continuum sensitivity is simulated with a power law model $E^{-\Gamma}$ with spectral index $\Gamma = 1$ and the energy range for the continuum source energy is taken from 0.1 MeV up to 10 MeV. All simulations are performed for an effective exposure time $T_{\text{eff}} = 10^{6}$ s, using the \textit{Cosima} module in MEGAlib. The physical interactions of simulated $\gamma$-ray photons within the payload are recorded for a certain angular acceptance window $\alpha_\text{acc}$ concerning the energy peak (for point sources) or energy bin (for continuum sources) of the source's energy spectrum. Moreover, the simulated sources are studied at the zenith, i.e. $(\theta,\phi) = (0,0)$.

Subsequently, simulated interactions are reconstructed and classified as tracked/untracked Compton or pair-production events based on the precision and details from the reconstruction algorithm using the \textit{Revan} module of MEGAlib and then analyzed with the \textit{Mimrec} module of MEGAlib, leading to the detector performance estimations.

The \textit{energy resolution} $R$ is an important parameter for the detector of any $\gamma$-ray telescope, representing the detector's capacity to measure the energy of incoming particles or radiation accurately. In this paper, the energy resolution is calculated as:
\begin{equation}
\label{ereseq}
    R = \frac{2.35\sigma}{E_0} \times 100\%,
\end{equation}
where $E_0$ is the energy of the source and the parameter $\sigma$ represents the standard deviation of the normal distribution fit to the peak of the source's energy spectrum. The energy spectrum was plotted within an acceptance window $w_\text{acc} = E_0 \pm 0.2E_0$ using 500 histogram bins.

The \textit{angular resolution measure} (ARM), for Compton telescopes, refers to the smallest angular distance between the direction of the incident $\gamma$-ray emitted by the source and each event \say{cone} traced out by the Compton-scattered $\gamma$-ray. It provides a measure of the MeV-CubeSat payload's ability to differentiate between the true direction of the incident photon and its measured direction. The ARM is defined by the following equation:
\begin{equation}
\label{armeq}
    \text{ARM} = \arccos{(\boldsymbol{\hat{e}}_g \cdot\boldsymbol{\hat{e}}_i)} - \varphi.
\end{equation}
where $\boldsymbol{\hat{e}}_g$ is the unit direction of the scattered gamma ray, $\boldsymbol{\hat{e}}_i$ the unit direction of the incident gamma ray, and $\varphi$ is the Compton scattering angle, defined in terms of the energies of the Compton scattered $\gamma$-ray ($E_g$) and electron ($E_e$) as:
\begin{equation}
\label{csa_eq}
    \cos{(\varphi}) = 1 - \frac{m_e c^2}{E_g} + \frac{m_e c^2}{E_g + E_e}.
\end{equation}
The ARM distribution represents the detector's point spread function (PSF). The narrower the PSF, the better the ARM. In this paper, the quantity of interest is $\sigma_\text{ARM}$, which is the Full Width at Half Maximum (FWHM) of the PSF peak within $\alpha_\text{acc}$. This quantity is calculated within an angular acceptance window $20^\circ \leq \alpha_\text{acc} \leq 50^\circ$, for both tracked and untracked Compton events. This limit for the angular acceptance window ensures the containment of the full ARM peak without excessive inclusion of the distribution's \say{tails}, therefore mitigating the majority of incompletely absorbed electrons or wrongly reconstructed events, keeping only the Compton scattered gamma rays with correct Compton kinematics.

\subsection{Continuum Sensitivity}
The \textit{continuum sensitivity}, denoted as $F_z$, is a measure of the CubeSat's ability to distinguish astrophysical sources from the background. Moreover, the continuum sensitivity is calculated within an energy window of $[0.1, 10]$ MeV, broken into $n+1$ logarithmic intervals for $n$ continuum sources. The sensitivity is calculated as:
\begin{equation}
\label{ContSens}
    F_z = \frac{z^2 + z\sqrt{z^2 + 4N_B}}{2T_\text{eff}A_\text{eff}},
\end{equation}
where $z = 3$ represents the significance, $A_\text{eff}$ is the effective area of the detector, and $N_B$ is the number of background photons within the angular resolution element defined by the detector. Furthermore, the effective area $A_\text{eff}$ is calculated as follows:
\begin{equation}
\label{aeffeq}
    A_\text{eff} = \frac{N_\text{sel}}{N_\text{gen}} \cdot \pi r^2,
\end{equation}
where $N_\text{sel}$ is the number of events within the ARM $\alpha_\text{acc}$ after event selection in \textit{Mimrec}, $N_\text{gen}$ is total number of generated events in \textit{Cosima}, and $r = 15$ cm is the radius of the \say{surrounding sphere}\footnotemark[4] \footnotetext[4]{See the Cosima manual for an explanation.}. The effective area can never become infinitely large even when $r \rightarrow \infty$, because $N_\text{sel} \rightarrow 0$ in this limit. Furthermore, the number of background photons is calculated as:
\begin{equation}
\label{Nb}
    N_B = \Phi \cdot T_\text{eff} \cdot A_\text{eff} \cdot \Delta\Omega.
\end{equation}
Here, $\Phi$ is the background flux determined from the background spectra in eq. \eqref{Bkgspectrum} as:
\begin{equation}
    \Phi = \int \sum_{i=1}^n \Big(\frac{dN}{dE}\Big)_i\, dE,
\end{equation}
and $\Delta\Omega$ is the angular resolution element:
\begin{equation}
    \Delta\Omega = [\cos(\overline{\varphi} - \sigma_\text{ARM}) - \cos(\overline{\varphi} + \sigma_\text{ARM})] \cdot 2\sigma_\text{SPD}.
\end{equation}
Here, $\overline{\varphi}$ is calculated by averaging Eq. \eqref{csa_eq}, and $2\sigma_\text{SPD}$ is calculated from the FWHM of the fitted scatter plane deviation (SPD) distribution for the scattered electron, which is expressed in terms of the unit direction of the initial $\gamma$-ray, the scattered $\gamma$-ray and the electron as follows:
\begin{equation}
\label{SPDcalc}
    \text{SPD} = \arccos{[(\boldsymbol{\hat{e}}_g \times \boldsymbol{\hat{e}}_i) \cdot (\boldsymbol{\hat{e}}_g \times \boldsymbol{\hat{e}}_e)]}.
\end{equation}
For untracked Compton events, $\boldsymbol{\hat{e}}_e$ is undefined, therefore $\sigma_\text{SPD}$ was set as $180^\circ$ for these events. The continuum sensitivity calculation in this study focuses on $\gamma$-ray sources located at high Galactic latitudes and observed at high zenith angles from the Earth.

Furthermore, in regards to a Compton telescope situated in a low-Earth orbit, the MeV CubeSat payload is ideally suited for a low-Earth orbit, approximately 550 km above the Earth's surface, characterized by a nearly equatorial path with an inclination of $\leq 5^{\circ}$) \cite{2022JCAP...08..013L}. This orbit is susceptible to radiation from instrumental and astrophysical sources. Hence, a comprehensive background study specific to a $\gamma$-ray telescope positioned in a low-Earth orbit is discussed below based on the background model provided in \cite{2019ExA....47..273C}.

\begin{description}
\item [Earth's Albedo Radiation:] Cosmic rays undergo scattering that generates secondary gamma rays upon interaction with Earth's atmosphere. These secondary gamma rays constitute a major component of the background radiation detected by space-based $\gamma$-ray telescopes \cite{2009PhRvD..80l2004A}. Although substantial emissions are not predicted, coincidental occurrences between Earth's albedo background and signal events in a Compton telescope at low angles are possible primarily due to the constraint that source reconstruction is confined to a great circle of celestial sphere.
    
\item [Extragalactic Background:] An isotropic background of gamma rays and neutrinos that permeates the whole Universe. This background comprises the extragalactic $\gamma$-ray background (EGRB) and the cosmic X-ray background (CXB). The origin of the extragalactic MeV background is still unknown \cite{2015ApJ...800L..27A, Ruiz-Lapuente_2016}, primarily due to the lack of observational data in this energy range. In addition to its role as a substantial constituent of the overall background, the EGB will assume a critical role as a subject of scientific inquiry for forthcoming missions centered on the MeV energy range \cite{2019BAAS...51c.290A}. 
    
\item [Charged Particle Background:] Cosmic rays, such as protons and atomic nuclei, come from outer space and have high energies. When they interact with the Earth's atmosphere, they create secondary particles like muons and $\gamma-$rays. These particles can reach the Earth's surface and contribute to background radiation. The energy of cosmic rays follows a power-law distribution, with intensity decreasing as energy increases. Solar flares and coronal mass ejections can produce bursts of high-energy particles contributing to background radiation. It is important to note that this study does not take into account the charged-particle background as the utilization of an anti-coincidence detector is capable of effectively vetoing a significant portion of this background \cite{Gargano:2021Q4}.
    
\item[Internal Material Activation:] Cosmic rays hitting the detector can cause ionization of the atoms inside the detector, creating radioactive isotopes. An equilibrium is eventually reached between new nuclei production and decay. As a result, a Compton telescope's instrumental background energy spectrum comprises a combination of continuous emission and characteristic lines originating from the decay of these radioactive isotopes. This spectrum partially overlaps with the signals recorded by the telescope. The overall background rate experienced by the telescope during its on-orbit operations relies on various factors, including the spacecraft's materials, payload composition, and orbital characteristics. Opting for an equatorial orbit is generally preferable to mitigate the impact of the South Atlantic Anomaly, a region known for elevated radiation levels.
\end{description}

\begin{figure*}[t]
    \begin{minipage}[c][0.65\textheight]{\textwidth}
    \centering
    \begin{subfigure}{0.485\textwidth}
        \includegraphics[width=\textwidth,trim={1cm 0.2cm 2.5cm 2.0cm}, clip]{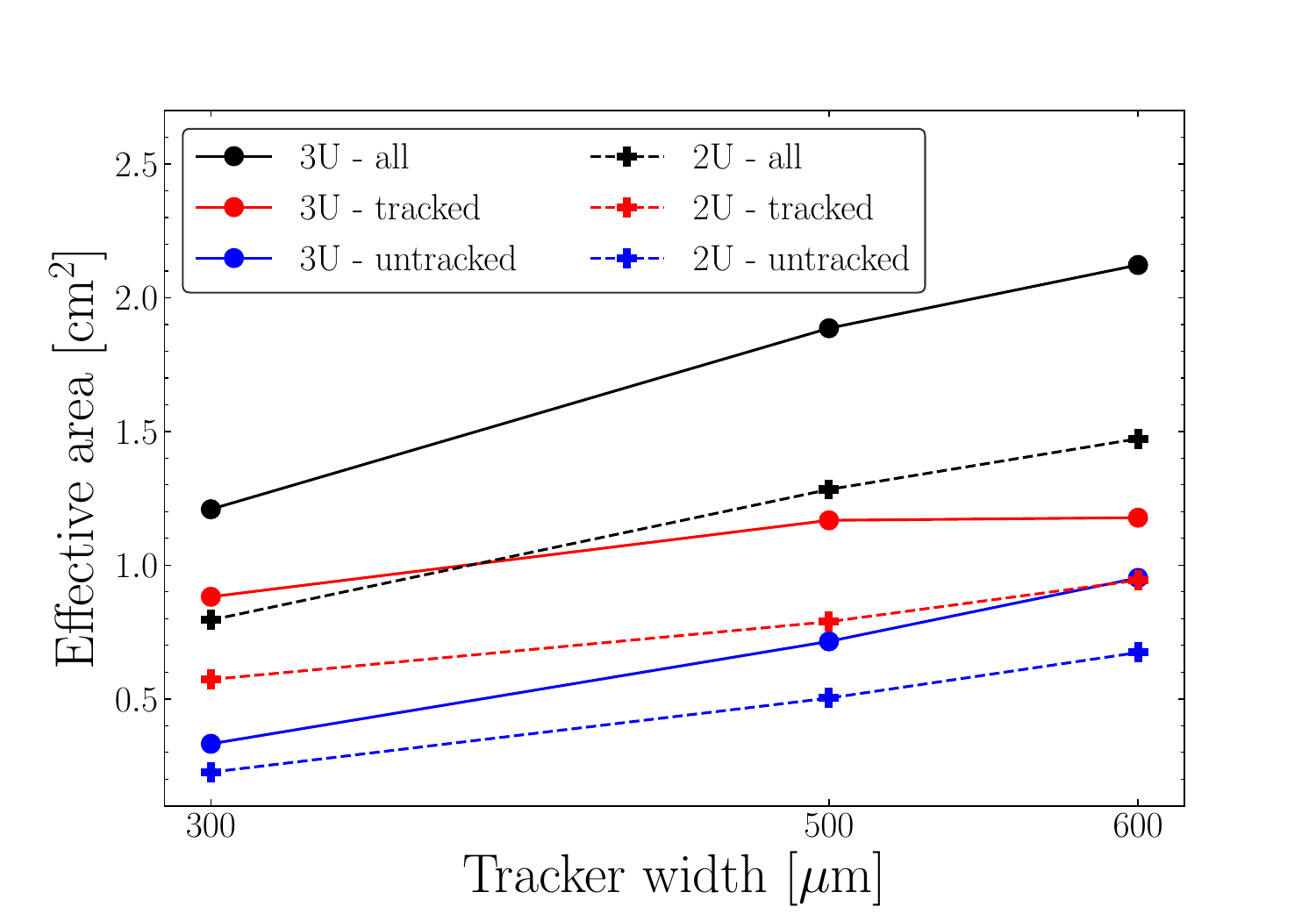}
        \caption{Effective area (Cal 0.5 cm)}
        \label{aeff_trsplit}
    \end{subfigure}
    \begin{subfigure}{0.485\textwidth}
        \includegraphics[width=\textwidth,trim={1cm 0.2cm 2.5cm 2.0cm}, clip]{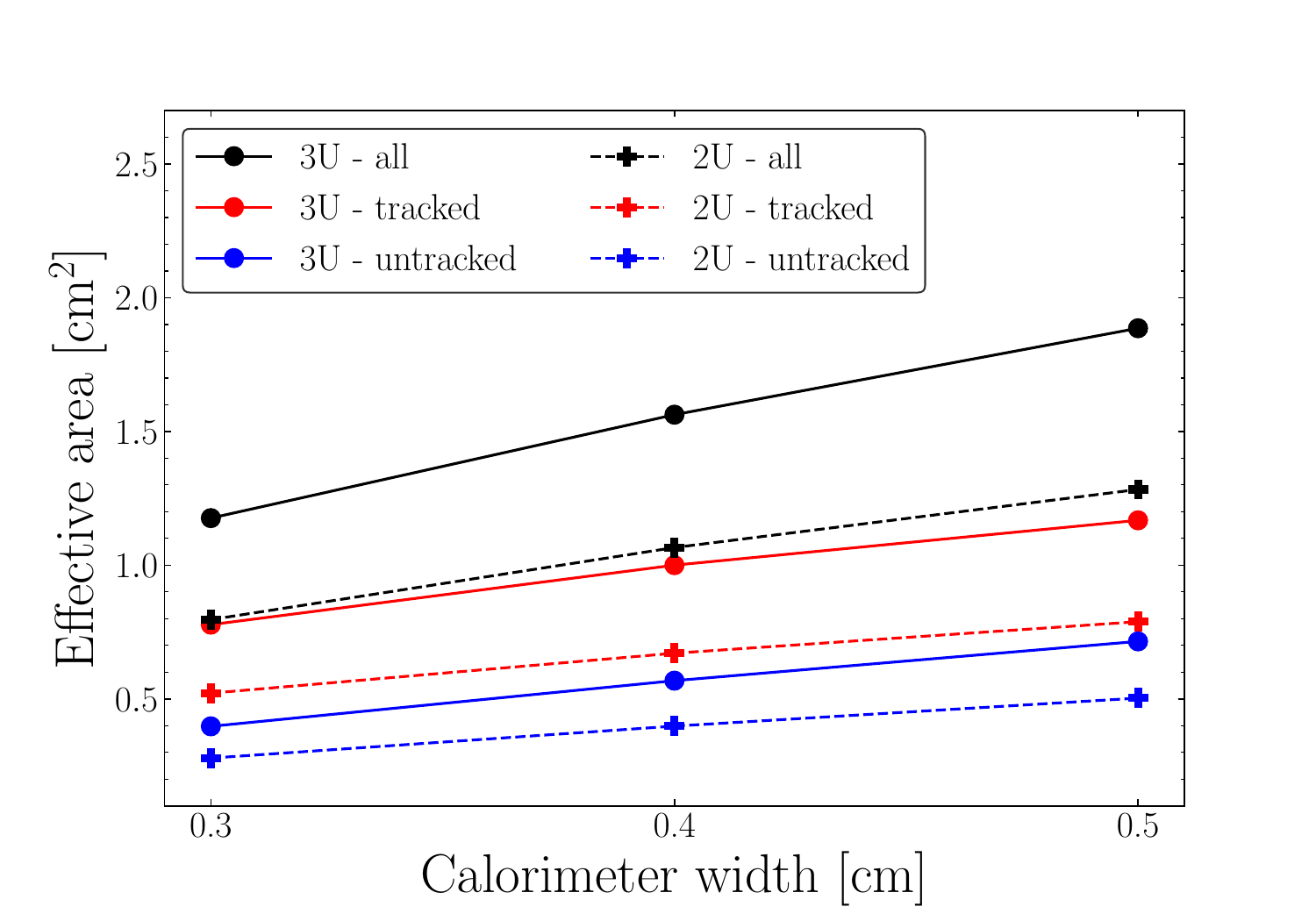}
        \caption{Effective area (Tr $500\ \mu$m)}
        \label{aeff_calsplit}
    \end{subfigure}
    \begin{subfigure}{0.485\textwidth}
        \includegraphics[width=\textwidth,trim={1cm 0.2cm 2.5cm 2.0cm}, clip]{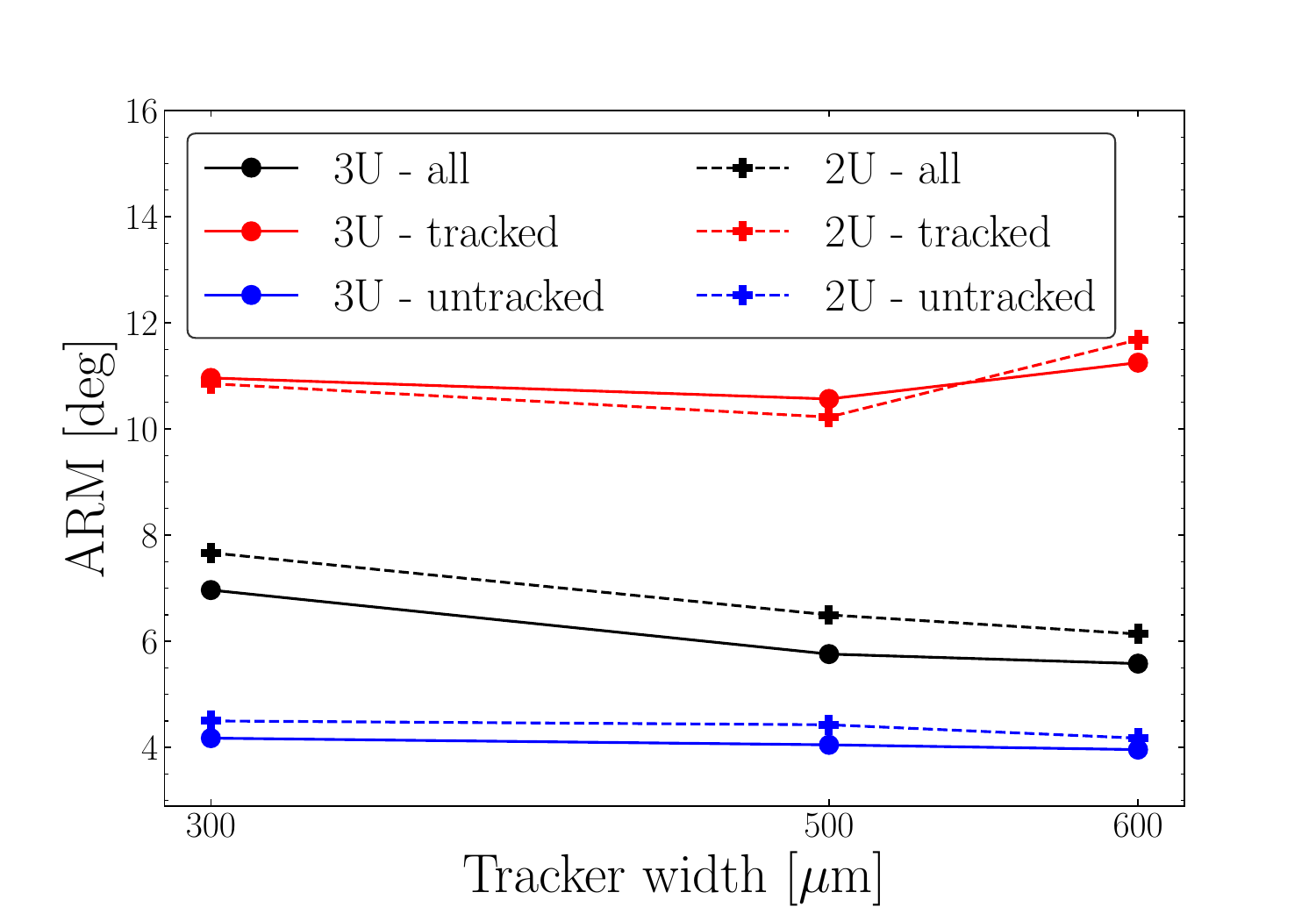}
        \caption{ARM (Cal 0.5 cm)}
        \label{arm_trsplit}
    \end{subfigure}
    \begin{subfigure}{0.485\textwidth}
        \includegraphics[width=\textwidth,trim={1cm 0.2cm 2.5cm 2.0cm}, clip]{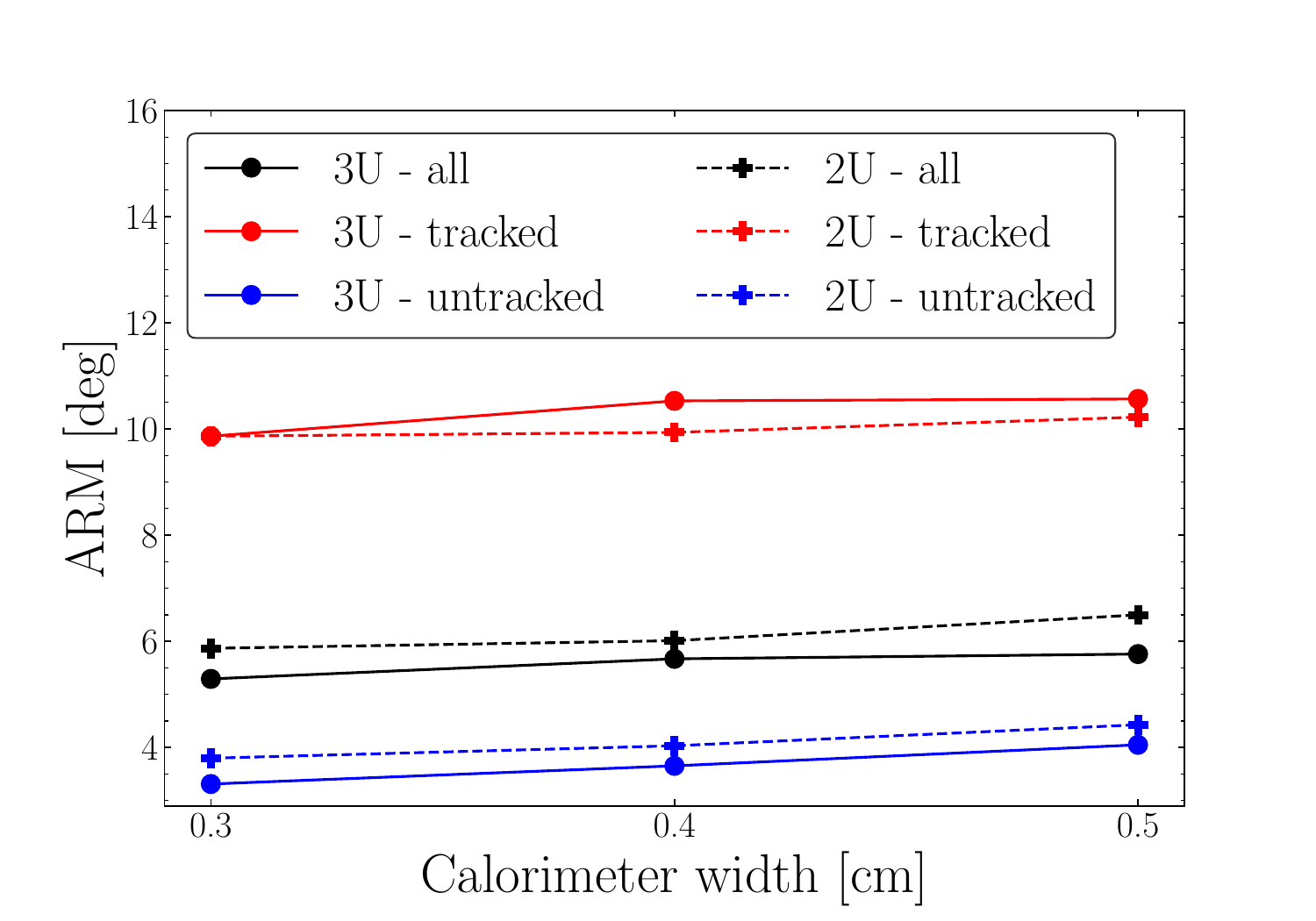}
        \caption{ARM (Tr $500\ \mu$m)}
        \label{arm_calsplit}
    \end{subfigure}
    \caption{Influence of varying tracker and calorimeter thicknesses on the effective area and angular resolution measure. Solid/dashed lines represent 3U/2U configurations.}
    \label{aeff_arm_trcsplit}
    \end{minipage}
\end{figure*}

\section{Results \& Discussion}
\label{results_section}
In this section, we present a comparison of various detector performance parameters across different tracker and calorimeter thicknesses, utilizing the geometry configuration outlined in \cite{2019AJ....158...42R} for the 2U and 3U payloads. In order to assess the detector performance, we calculate the Effective Area, ARM, Energy Resolution, and Continuum Sensitivity for the different simulated detector designs at 1 MeV. Given the dominance of Compton interactions in the energy range considered, we further divided the effective area, ARM, energy resolution, and continuum sensitivity results into a \textit{tracked}, \textit{untracked} and \textit{all} category, where \say{tracked} and \say{untracked} interactions denote those with and without a measurable scattered electron track, while \say{all} refers to both tracked and untracked interactions. 

\subsection{Payload design performance}
In Figure~\ref{aeff_arm_trcsplit}, we present a comparison of the CubeSat's effective area and ARM, varying the default widths of the calorimeter (Cal) and tracker (Tr) for both the 2U and 3U payloads.
When adjusting tracker and calorimeter thicknesses, clear trends are observed in both the effective area and ARM. Figure~\ref{aeff_trsplit} illustrates that increasing the tracker width leads to a larger effective area, as the increased Compton cross-section enhances particle detection and reconstruction. Meanwhile, Figure~\ref{aeff_calsplit} demonstrates that increasing the calorimeter thickness similarly results in increased effective area.

\begin{figure*}[t]
    \begin{minipage}[c][0.65\textheight]{\textwidth}
    \centering
    \begin{subfigure}{0.485\textwidth}
        \includegraphics[width=\textwidth,trim={1cm 0.2cm 2.5cm 2.0cm}, clip]{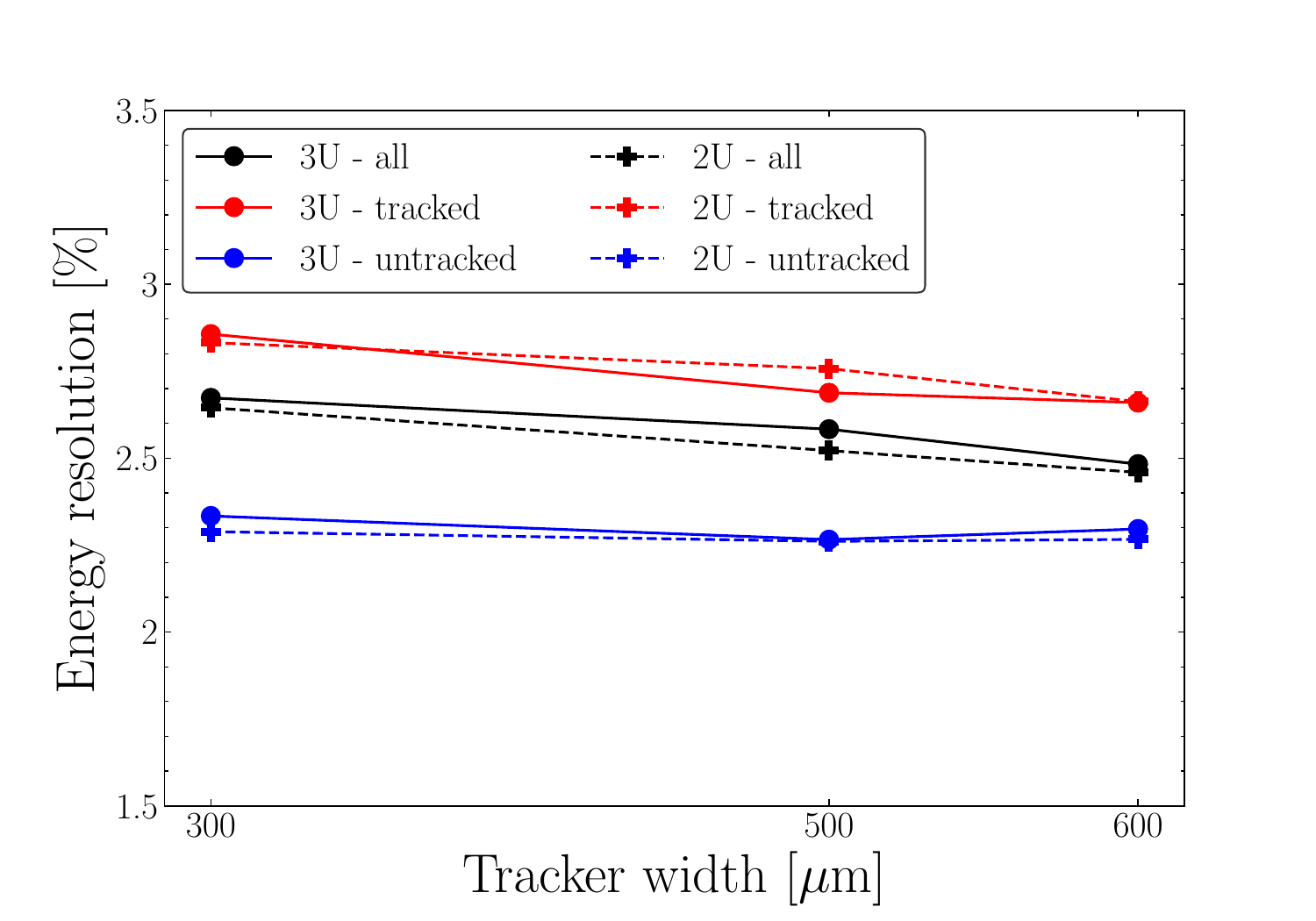}
        \caption{Energy resolution (Cal 0.5 cm)}
        \label{eres_trsplit}
    \end{subfigure}
    \begin{subfigure}{0.485\textwidth}
        \includegraphics[width=\textwidth,trim={1cm 0.2cm 2.5cm 2.0cm}, clip]{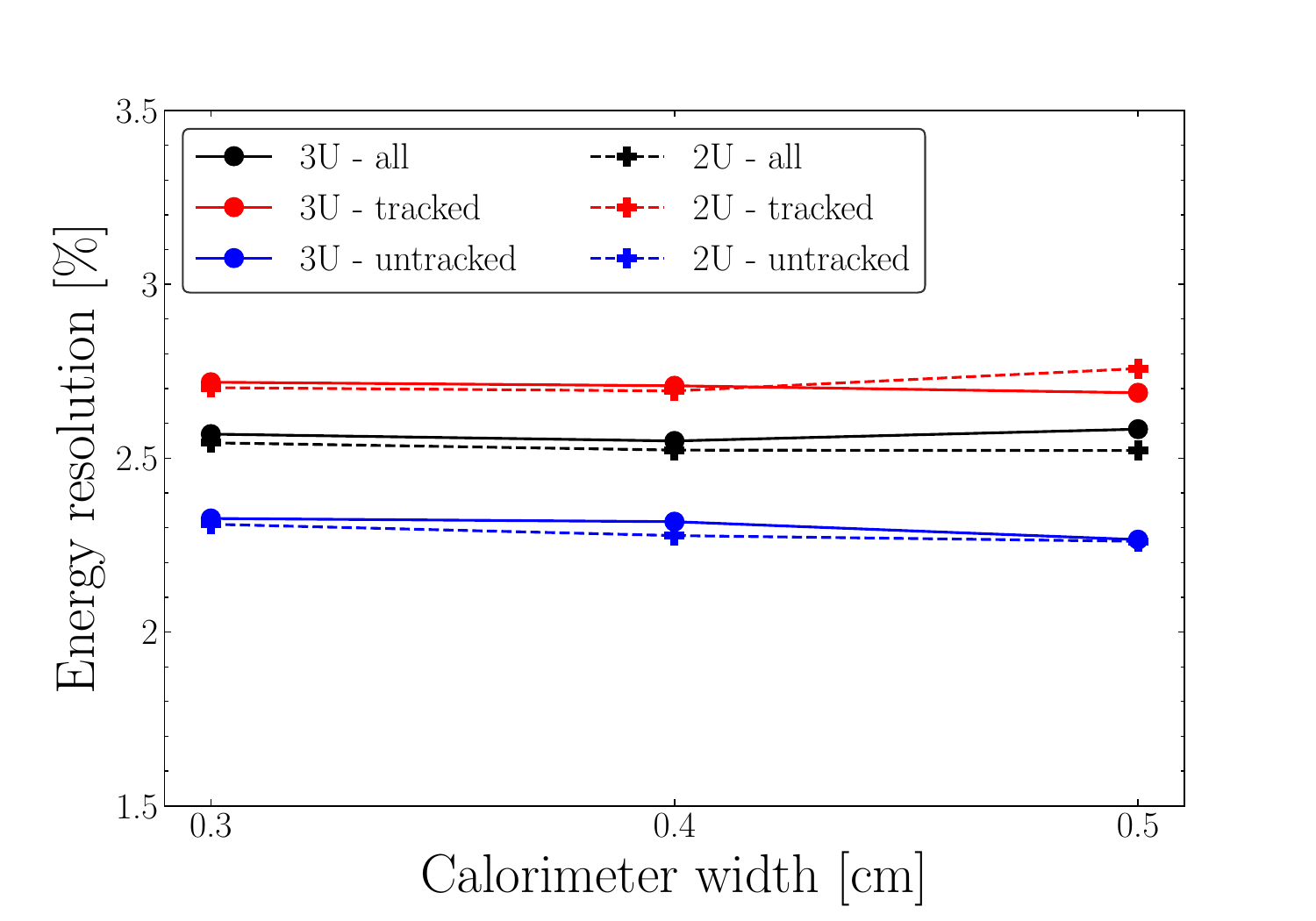}
        \caption{Energy resolution (Tr $500\ \mu$m)}
        \label{eres_calsplit}
    \end{subfigure}
    \begin{subfigure}{0.485\textwidth}
        \includegraphics[width=\textwidth,trim={1cm 0.2cm 2.5cm 2.0cm}, clip]{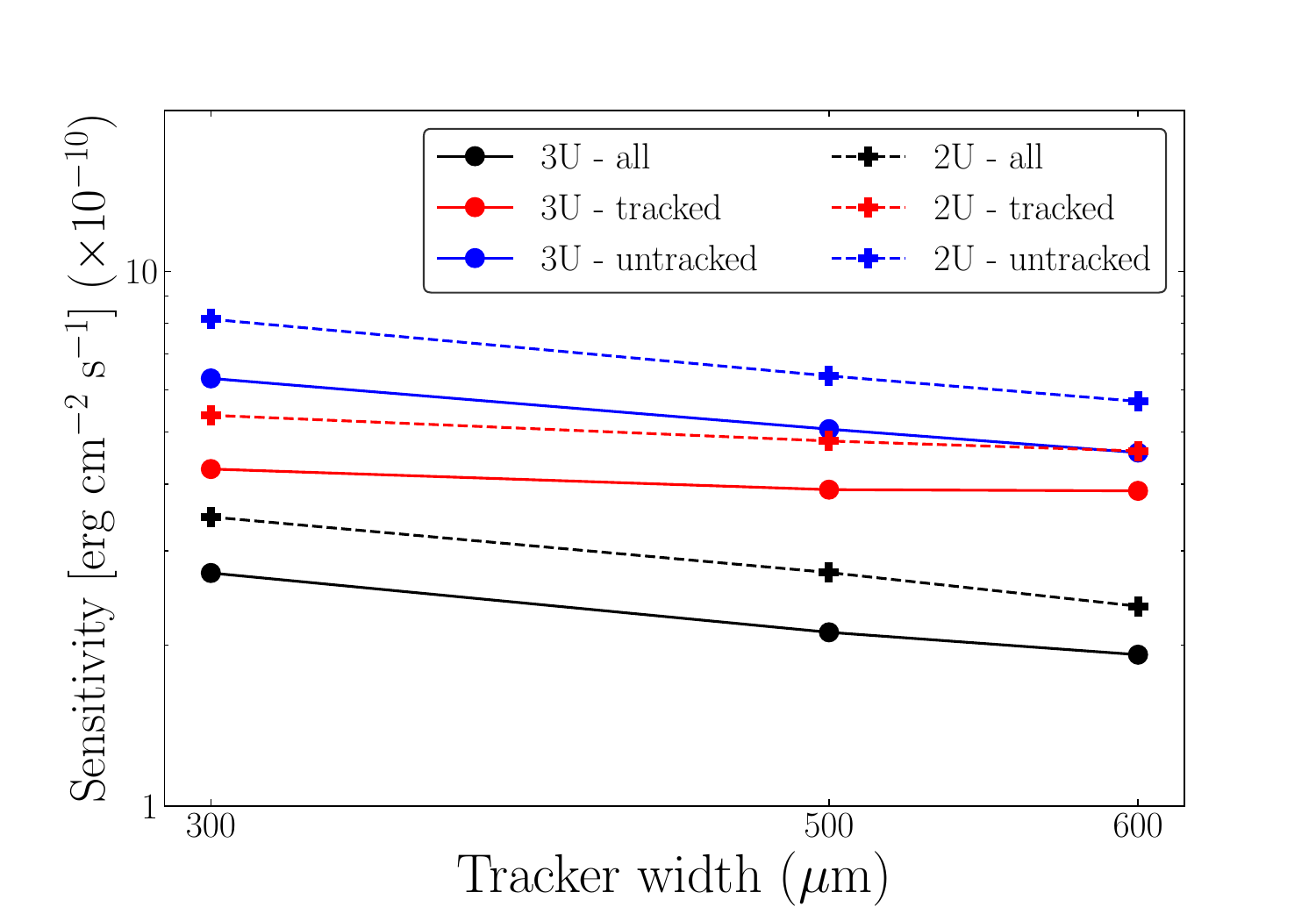}
        \caption{Sensitivity (Cal 0.5 cm)}
        \label{sensi_trsplit}
    \end{subfigure}
    \begin{subfigure}{0.485\textwidth}
        \includegraphics[width=\textwidth,trim={1cm 0.2cm 2.5cm 2.0cm}, clip]{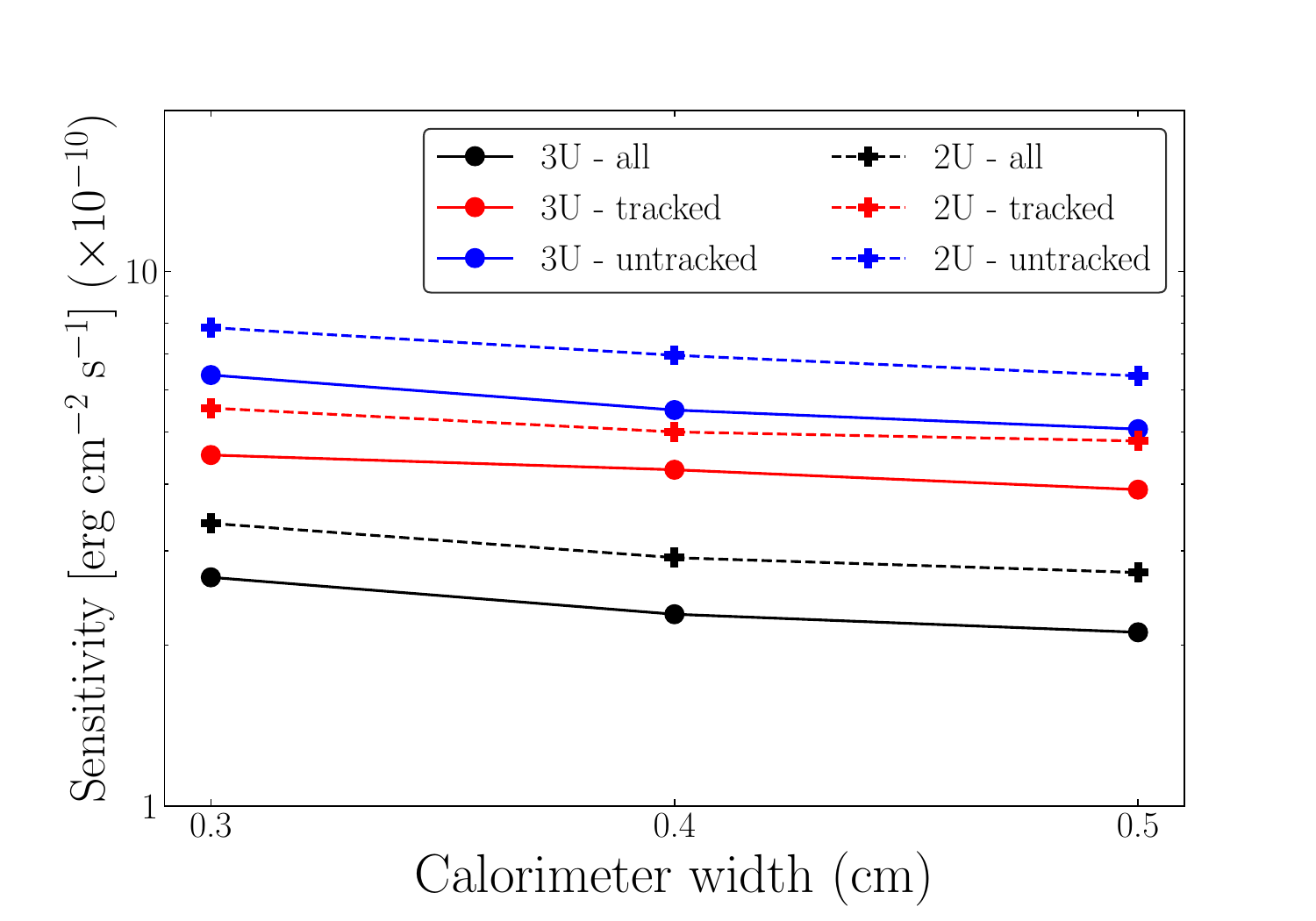}
        \caption{Sensitivity (Tr $500\ \mu$m)}
        \label{sensi_calsplit}
    \end{subfigure}
    \caption{Influence of varying tracker and calorimeter thicknesses on the energy resolution and continuum sensitivity. Solid/dashed lines represent 3U/2U configurations.}
    \label{eres_sensi_trcsplit}
    \end{minipage}
\end{figure*}

As shown in Figure~\ref{arm_trsplit}, increasing the tracker thickness enhances the ARM, reflecting the more precise tracking of Compton events. In contrast, increasing the calorimeter thickness decreases the ARM as shown in Figure~\ref{arm_calsplit}, given the inverse relationship $\text{ARM} \propto 1/E$. While the effective area for \textit{untracked} events within $\alpha_\text{acc}$ is smaller compared to the \textit{tracked} events due to the lower number of reconstructed \textit{untracked} events, \textit{untracked} events exhibit superior ARM performance. Primarily due to the absence of electron tracking, eliminating \say{sidelobes} caused by incompletely absorbed electrons and resulting in a more focused ARM peak.

Figure~\ref{eres_sensi_trcsplit} presents a comparative analysis of energy resolution and sensitivity across different payload designs. In terms of energy resolution, variations in tracker and calorimeter thicknesses (Figures \ref{eres_trsplit} and \ref{eres_calsplit}) have a negligible effect since energy resolution is the property of the material. However, when considering tracked/untracked events, the energy resolution is impacted. This is because, in the case of tracked interactions, the energy of the scattered electrons is taken into account, leading to a broadening of the energy spectrum peak and, consequently, a poorer energy resolution.
The trend in sensitivity for tracked and untracked interactions is found to decline, as shown in Figures~\ref{sensi_trsplit} and \ref{sensi_calsplit}. Untracked interactions result in an inferior sensitivity because the SPD FWHM $=360^\circ$ for these interactions, leading to a significantly higher number of background photons compared to tracked interactions. The sensitivity for \textit{all} interactions is best because the effective area is largest in that case, and the SPD FWHM is the same for tracked interactions.

During the simulation of MeV $\gamma$-rays, various high-energy particles are generated as a result of interactions between the $\gamma$-rays and the payload. The MeV $\gamma$-rays interact with the payload mostly through Compton scattering, yielding Compton events. In addition to Compton interactions, there are other high-energy processes, such as pair production, that affect the simulation performance. 

\begin{figure*}[t]
    \centering
    \includegraphics[width=\textwidth]{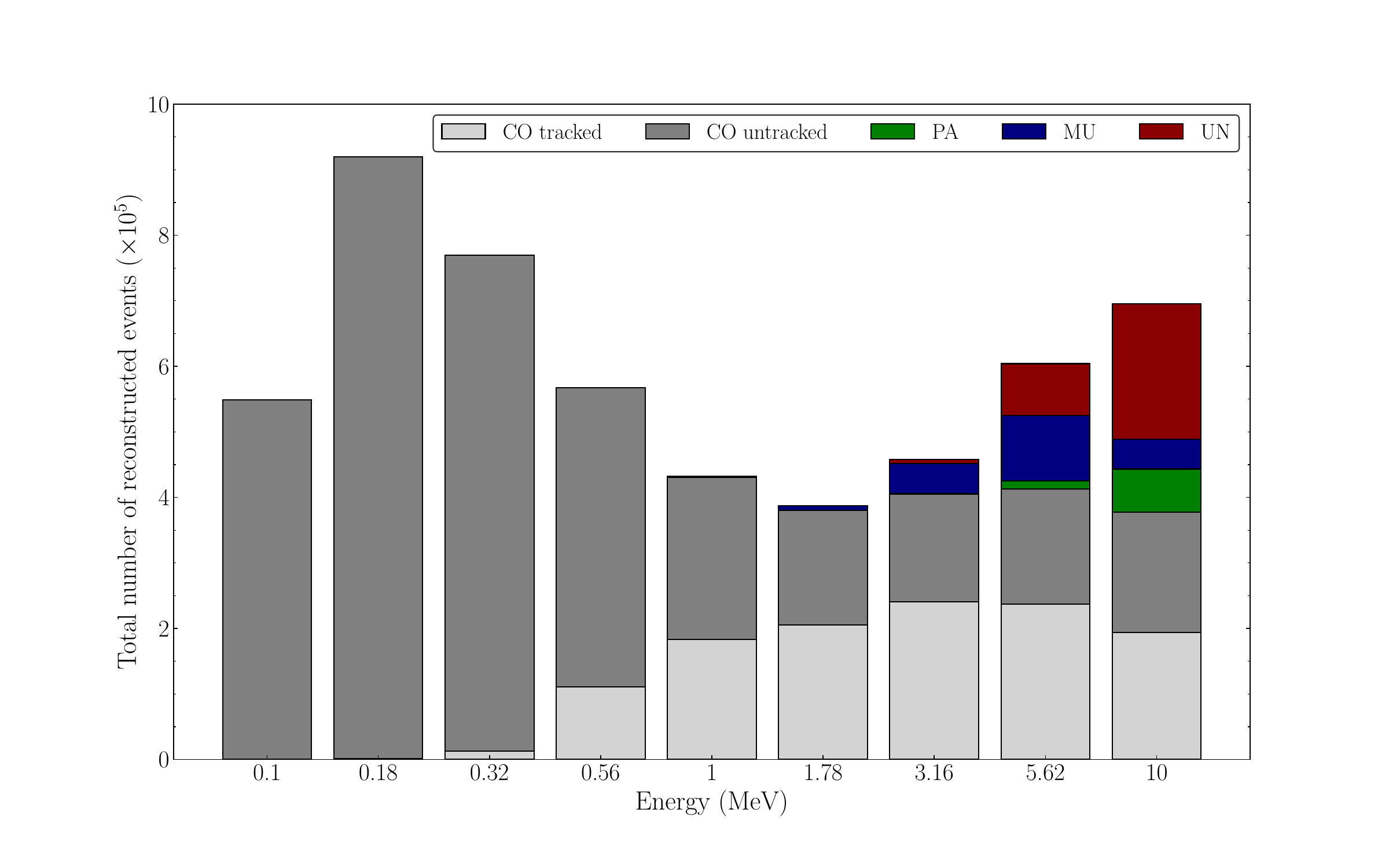} 
    \caption{Distribution of reconstructed event type classes for a 3U configuration.}
    \label{3u_barplot}
\end{figure*}

Figure~\ref{3u_barplot} shows the total number of reconstructed events and types of events over the detector energy range for a 3U payload configuration. From the figure, we note that tracked Compton events (i.e., the events that contain information about the position vectors of both the scattered $\gamma$-ray and electron) start to occur above 0.1 MeV. Since there are no tracked Compton events at energies below 0.1 MeV, a measure for $\boldsymbol{\hat{e}}_e$ does not exist. The maximum number of Compton interactions are observed at $\sim$0.18 MeV, which significantly contributes to the local maxima of effective area at this energy.

\begin{figure*}[t]
    \centering
    \begin{subfigure}{0.485\textwidth}
        \includegraphics[width=\textwidth,trim={0.1cm 0.2cm 2.5cm 1.7cm}, clip]{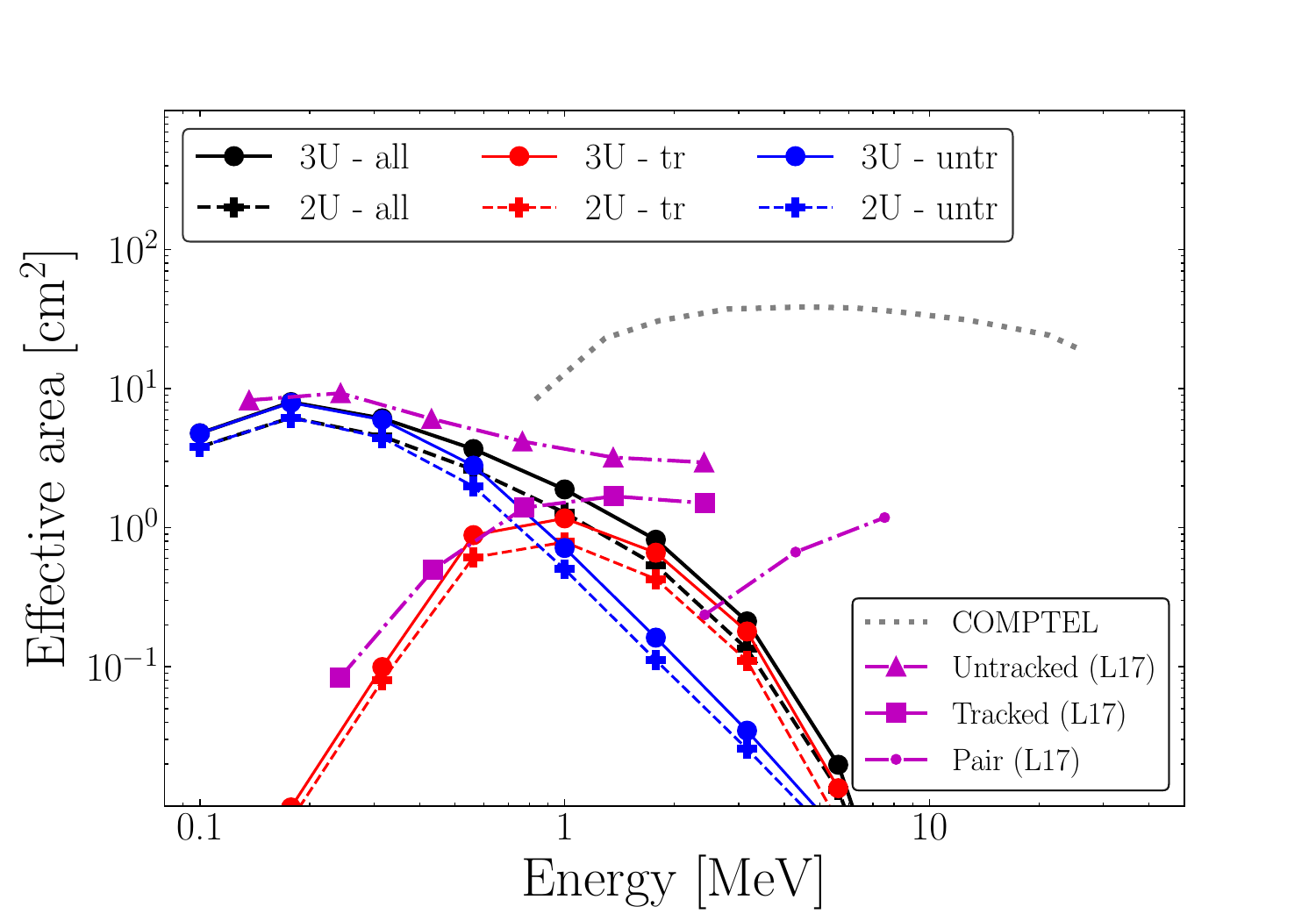}
        \caption{Effective area}
        \label{aeff_split}
    \end{subfigure}
    \begin{subfigure}{0.485\textwidth}
        \includegraphics[width=\textwidth,trim={0.1cm 0.2cm 2.2cm 1.7cm}, clip]{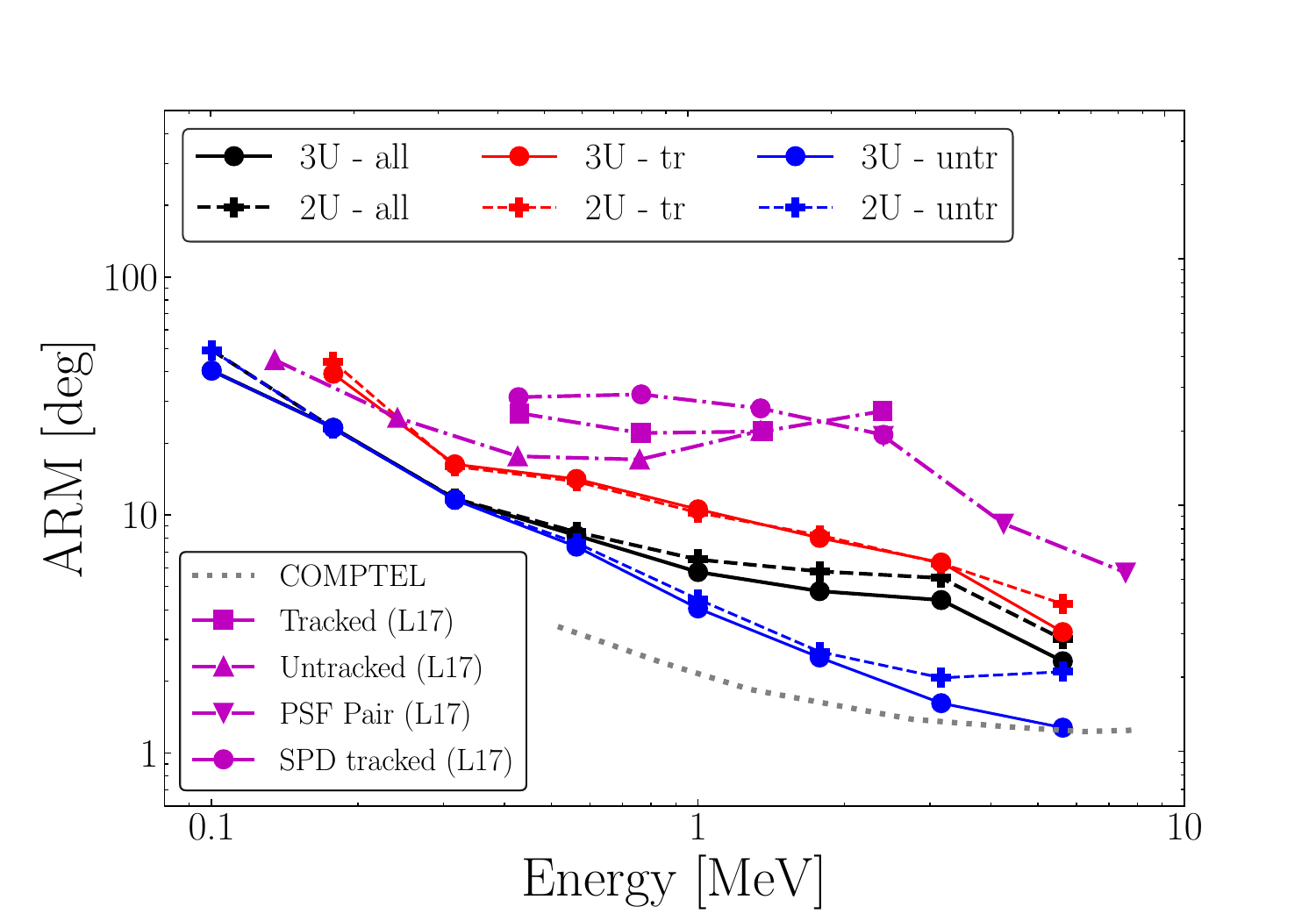}
        \caption{ARM}
        \label{arm_split}
    \end{subfigure}
    \caption{Energy dependent simulated effective area and ARM comparison of \textit{Tr:} $500\ \mu$m and \textit{Cal:} $0.5$ cm with other detector results. L17 data points are adopted from (\cite{2017AJ....153..237L}). Other data points for instruments such as COMPTEL \cite{Schonfelder1993}. \textit{Tracked} events are shown in red, and \textit{untracked} events are shown in blue.}
    \label{Tr_untr_L17}
\end{figure*}

Furthermore, we note that for the energy range below 4 MeV, Compton interactions dominate, whereas, above 4 MeV, pair production starts to become relevant; hence, the $\gamma-$ray photons above 4 MeV (mainly between 4-30 MeV) cause a significant \say{mixture} of both Compton and pair-production events. This mixture leads to event reconstruction problems, especially at high energies (e.g., $>$10 MeV). Moreover, at the comparatively high-energy range, the reconstructed events need to be labelled accurately, and as a result, a large percentage of them fall under the category of unknown (UN) events. These UN events occur because, after 4 MeV, there is a significant possibility that an incident photon causes Compton scattering in the first (few) tracker layer(s). However, afterwards, the Compton-scattered gamma ray(s) still has/have enough energy to initiate pair production. In such a case, MEGAlib cannot correctly classify the event, marking it as a UN event. Because of the large number of UN events, fewer Compton events can be used to estimate the detector's performance parameters, leading to uncertain values. In order to reduce the percentage of UN events and improve the CubeSat detector's performance accuracy at high energies, new techniques are required e.g. incorporating machine learning algorithms.

\subsection{Comparative study of simulated performance}
The simulated performance parameters are calculated and compared among different payload designs and interaction types and with other detectors (i.e., COMPTEL, Integral, etc.). 

In Figure~\ref{Tr_untr_L17}, performance calculations for \textit{all} and \textit{(un)tracked} interactions for the default payload configuration within the MeV gap are compared with \cite{2017AJ....153..237L}. 
Figure~\ref{aeff_split} shows and compares the simulated effective area for (un)tracked and combined. This simulated payload only concentrates on Compton events. Thus, the reconstructed pair events are negligible in this case. In the case of (un)tracked events, a similar effective area is reached up to 1 MeV.
Figure~\ref{arm_split} shows and compares the simulated ARM for (un)tracked and combined events. Here, the CubeSat performs better in terms of ARM across all energies, 

\begin{figure*}[t]
    \centering
    \includegraphics[width=\textwidth]{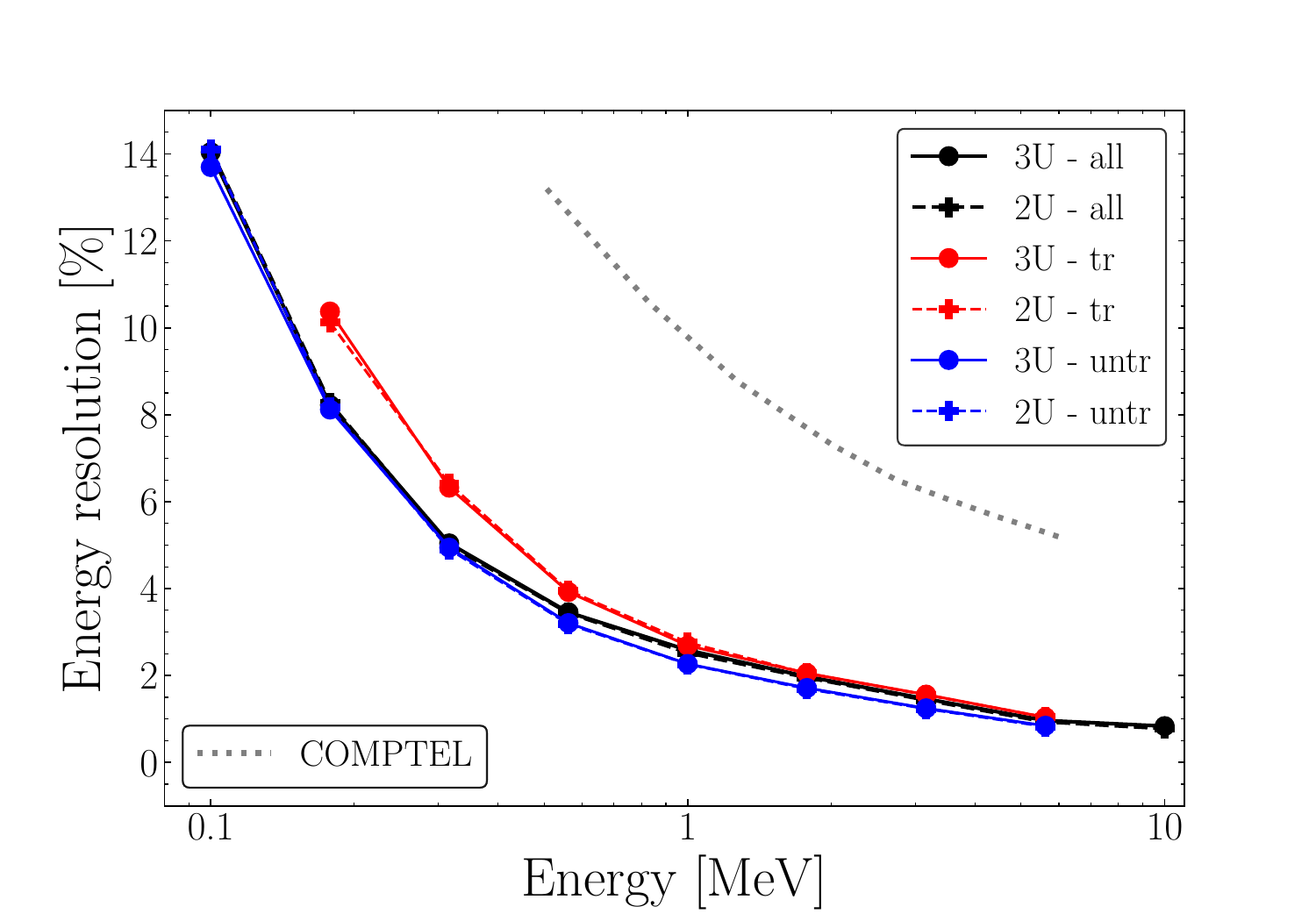} 
    \caption{Energy dependent simulated energy resolution  comparison of $Tr: 500\ \mu$m and $cal: 0.5$ cm with other detector results. The data points for COMPTEL are taken from \cite{Schonfelder1993}.}
    \label{eres_compare}
\end{figure*}

Figure~\ref{eres_compare} shows a comparison of energy resolution for simulated (un)tracked and combined events. This parameter remained the same for different payload sizes for the same type of events, further indicating that energy resolution is a property of the payload material and independent of the payload design. This parameter is compared with COMPTEL, and the proposed MeV CubeSat achieves much better energy resolution than COMPTEL due to technological advancements in gamma-ray detectors. 
 
\begin{figure*}[t]
    \centering
    \includegraphics[width=\textwidth]{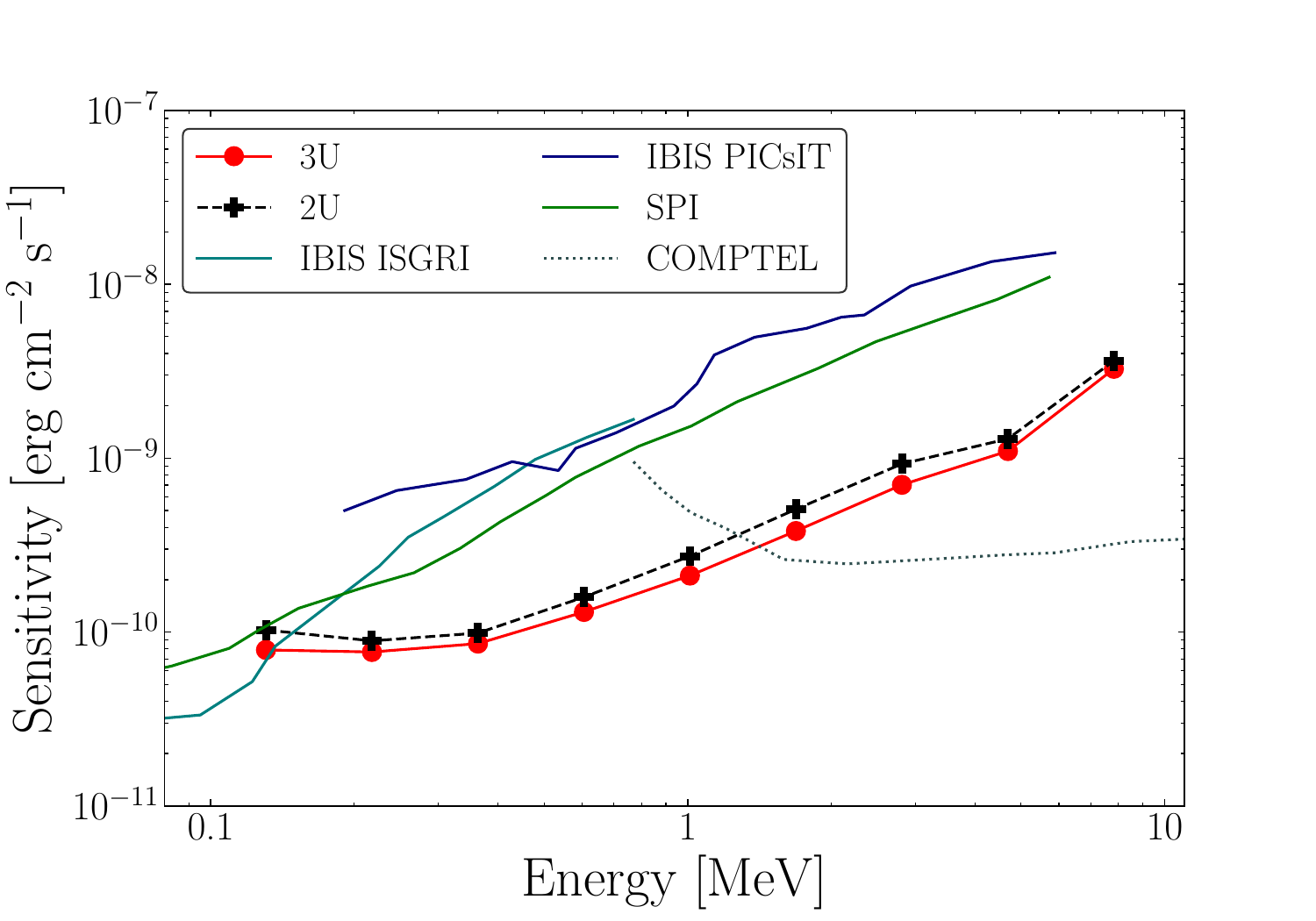} 
    \caption{Default 2U and 3U configurations continuum sensitivity within the MeV gap compared with literature. Past/current/future missions are shown as dotted/solid/dashed lines. L17 data points are adopted from (\cite{2017AJ....153..237L}). Other data points for instruments such as COMPTEL \cite{Schonfelder1993}, SPI \cite{SPIref}, and IBIS \cite{IBISref} are also plotted.}
    \label{Sensi_compare}
\end{figure*}

In Figure~\ref{Sensi_compare}, the continuum sensitivity is compared to COMPTEL and INTEGRAL. While it can be noted that the MeV CubeSat payload achieves better sensitivity compared to COMPTEL only up to around 1 MeV, it achieves similar/better sensitivity across the whole MeV gap (spanning from $\sim 10^{-10}$ to $\sim 10^{-9}$\fzu) compared to IBIS and SPI and could therefore serve as a replacement for these instruments. 

\begin{figure*}[t]
    \centering 
    \includegraphics[width=0.9\textwidth]{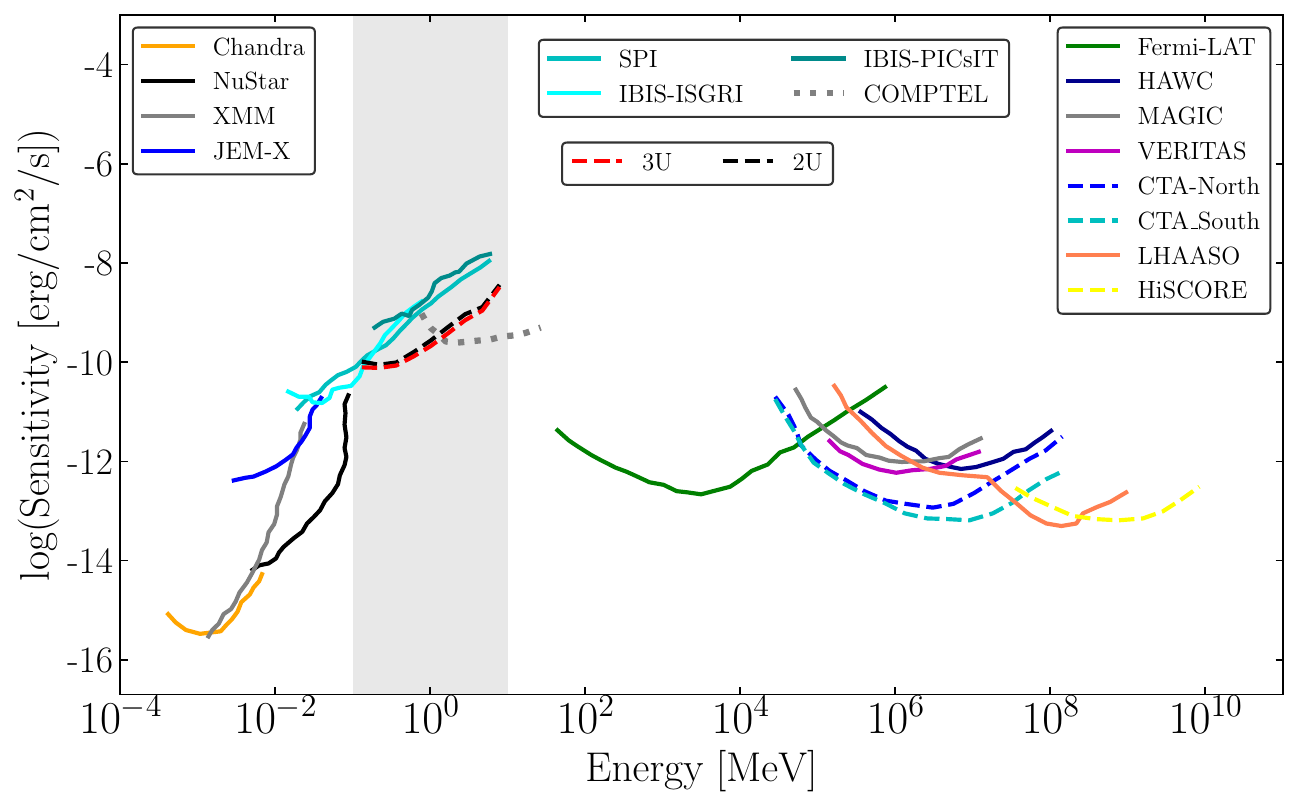}
    \caption{Comparison among sensitivities of active instruments with the proposed 2U and 3U detector design. Other instruments sensitivities are adopted from \cite{2018JHEAp..19....1D}. Past/current/future missions are shown in dotted/solid/dashed lines, respectively. The MeV gap is shown as a grey band.}
    \label{sensi}
\end{figure*}

A more complete comparison of the CubeSat detector's sensitivity with other instruments is shown in Figure~\ref{sensi}. Here, the active missions are plotted with solid lines. Past missions (e.g., COMPTEL) are marked with dotted lines, and future missions are plotted with dashed lines. 

\section{Summary and conclusions}
\label{conclusion}

Based on the simulations, we carried out a comprehensive comparison between various payload configurations, each featuring varying tracker and calorimeter specifications. We find that the effective area and ARM depend on the thickness of the tracker and calorimeter. However, the relationship between energy resolution and continuum sensitivity with regard to the tracker and calorimeter thickness remains unclear. The study also concluded that the effective area and ARM are directly proportional to the payload size. In contrast, the energy resolution and continuum sensitivity show no discernible correlation with the payload size. Taking into account the simulation study and the constraints imposed by electronics and cost, we have determined that a tracker thickness of 500 $\mu$m and a calorimeter thickness of 0.5 cm constitute the most suitable configuration for our payload. 

Moreover, MeV $\gamma$-ray payloads contained within CubeSat dimensions represent a significant advancement toward understanding the viability of current technologies for larger-scale satellite missions in the MeV energy range. Regardless of the chosen gamma-ray detection technology, the data collected from CubeSats can be valuable for calibrating upcoming missions. Additionally, MeV payloads utilizing CubeSat technologies represent a cost-effective way to pioneer future space missions and emerging space technologies.

\section*{Acknowledgments}
This research is supported by HKU-RMGS Funds (207300301,207301033; P.I.: Prof.~Q.~A.~Parker).
The research of P.~S.~Pal is partially supported by a General Research Fund (GRF) grant from the Research Grants Council of the Hong Kong Special Administrative Region, China (HKU Project 17304920; P.I. Dr.~Stephen C.~Y.~Ng). We acknowledge Prof. Riccardo Rando for his support of detector geometry.

\subsection*{Author Contributions} 
RD, KdK, AK, PSPal contributed to the simulation, data analysis, plots, and manuscript writing. PSParkinson, AR, QAP supervised the project and revised the manuscript.

\subsection*{Funding}
This research work is supported by HKU-RMGS funds (207300301,207301033; P.I.: Prof.~Q.~A.~Parker). PSPal is partially supported by the HKU-GRF fund (17304920; P.I. Dr.~Stephen C.~Y.~Ng). 

\subsection*{Conflicts of Interest}
There is no conflict of interest in this research work. 

\subsection*{Data Availability}
The simulation software MEGAlib(\url{https://megalibtoolkit.com/setup.html}) is publicly available on github(\url{https://github.com/zoglauer/megalib}). We acknowledge Prof. Riccardo Rando for his concept design of detector geometry and background model. Anybody can reproduce the results in this paper. 

\section*{Supplementary Materials}
N/A

\bibliographystyle{plain}
\bibliography{test.bib}

\end{document}